\documentclass[letterpaper,twocolumn,10pt]{article}
\usepackage{usenix2019_v3}

\usepackage{tikz}
\usepackage{amsmath}

\usepackage{filecontents}

\usepackage{microtype}
\usepackage{url}

\usepackage{subfig}
\usepackage{multirow}
\usepackage{booktabs}
\usepackage{graphicx}
\usepackage{xspace}
\usepackage{algorithm}
\usepackage[noend]{algpseudocode}
\usepackage{balance}
\usepackage{mathtools}
\usepackage{amsfonts}
\usepackage{amsthm}
\usepackage{threeparttable}
\usepackage{xcolor}
\usepackage{enumitem,kantlipsum}
\usepackage{siunitx}
\usepackage[compact]{titlesec}
\usepackage[nosort]{cleveref} 
\usepackage{url}
\usepackage{subcaption}
\usepackage{subfig}
\usepackage{amssymb}
\usepackage{multicol}
\usepackage{multirow}
\usepackage{wrapfig}
\usepackage{xspace}
\usepackage{pifont}
\newcommand{\cmark}{\ding{51}}%
\newcommand{\xmark}{\ding{55}}%
\newcommand{\pmark}{\ding{115}}%

\NewDocumentCommand{\var}{O{s} m O{}}{%
  \ensuremath{#1_{#2}^{#3}}
}
\usepackage{siunitx}
\usepackage{colortbl}

\newcommand{\commentout}[1]{}

\definecolor{light-gray}{gray}{0.80}

\newcommand\fref{Fig.~\ref}

\newcommand\sref{\S~\ref}

\usepackage{amsthm}

\usepackage{amsthm}

\newtheorem{mytheorem}{Theorem}
\newtheorem{assumption}[mytheorem]{Assumption}
\newtheorem{lemma}[mytheorem]{Lemma}
\newtheorem{remk}[mytheorem]{Remark}

\newcommand{\UCP}{UCP\xspace}

\newcommand{\uniquepattern}{\emph{Unique}\xspace}
\newcommand{\replicatepattern}{\emph{Replicate}\xspace}
\newcommand{\shardpattern}{\emph{Shard}\xspace}
\newcommand{\shardpatternv}{\emph{Shard-V}\xspace}
\newcommand{\shardpatternh}{\emph{Shard-H}\xspace}
\newcommand{\partialpattern}{\emph{Partial}\xspace}

\newcommand{\padfree}{\emph{StripPad}\xspace}
\newcommand{\union}{\emph{Union}\xspace}
\newcommand{\extract}{\emph{Extract}\xspace}
\newcommand{\genmeta}{\emph{UcpInfo}\xspace}
\newcommand{\load}{\emph{Load}\xspace}
\newcommand{\save}{\emph{Save}\xspace}

\newcommand{\pp}{pipeline parallelism\xspace}
\newcommand{\tp}{tensor parallelism\xspace}
\newcommand{\Pipe}{Pipeline}

\newcommand{\tunespace}[1]{}

\newcommand{\hide}[1]{}

\newcommand{\revise}[1]{{\color{black}#1}}
\begin{document}

\date{}


\title{\Large \bf Universal Checkpointing: A Flexible and Efficient Distributed Checkpointing System for Large-Scale DNN Training with Reconfigurable Parallelism}

\author{
{\rm Xinyu Lian}\\
UIUC
\and
{\rm Sam Ade Jacobs}\\
Microsoft
\and
{\rm Lev Kurilenko}\\
Microsoft
\and
{\rm Masahiro Tanaka}\\
Microsoft
\and
{\rm Stas Bekman}\\
Snowflake
\and
{\rm Olatunji Ruwase}\\
Microsoft
\and
{\rm Minjia Zhang}\\
UIUC
} 

\maketitle

\begin{abstract}
Deep neural network (DNN) training continues to scale rapidly in terms of model size, data volume, and sequence length, to the point where multiple machines are required to fit large models for training. Different distributed and parallel training strategies have been developed to support large-scale DNN training by partitioning the training state across GPUs. However, existing DNN training systems provide very limited support for reconfiguring parallelism strategies in the middle of the training via checkpointing. This limitation arises because distributed checkpoints are tightly coupled to specific model parallelism and hardware configurations, preventing large-scale training jobs from efficiently adapting to hardware failures or resource elasticity. 

This paper presents \emph{Universal Checkpointing} (\UCP), a novel checkpointing system that enables flexible and efficient DNN training with reconfigurable parallelism. \UCP overcomes challenges in existing systems by decoupling checkpoint structure from parallel training strategies and hardware configurations. In addition, we present a pattern-based reconfiguration pipeline that enables automatic, flexible, and efficient mapping of checkpoint state to various parallelism strategies. Evaluation on a range of DNN models, including state-of-the-art dense and sparse LLMs, shows that \UCP enables reconfiguration for a broader set of widely used parallelism strategies than existing solutions while adding negligible reconfiguration cost. \UCP has been successfully employed in real LLM training workloads, greatly enhancing their flexibility and resilience to dynamic hardware environments. 

\end{abstract}

\section{Introduction}
\label{sec:intro}

The emergence of Large Language Models (LLMs)~\cite{foundation-models} has opened up new opportunities across various fields. In comparison to traditional methods, LLMs such as ChatGPT~\cite{chatgpt} and GPT-4~\cite{gpt4} have exhibited unique capabilities such as instruction following, commonsense reasoning, and few-shot generalization~\cite{scaleing-law-nlp}. These abilities have been achieved primarily through unsupervised learning on massive model sizes, ranging from billions to even trillion-scale parameters~\cite{llama3TechReport,mistral-7b,xai,phi3,pathway}).
However, as DNN continues to scale in model size, data size, and sequence length, multiple machines are required to fit the model for training. This has been made possible primarily through advanced distributed systems and parallelism technologies, where a model is partitioned across multiple devices and leverage the aggregated memory and compute capability from many GPU devices to train large-scale models~\cite{megatron-lm,gpipe,1f1b,deepspeed-ulysses,megatron-sp,zero-optimizer,megatron-lm-v2,mt530b}. 

While demonstrating excellent training efficiency and scalability improvement, training modern DNNs/LLMs can still span several days or even months on highly optimized GPU clusters~\cite{googlegemini,claude}. For example, GPT-4 is reportedly trained on \textasciitilde25,000 NVidia A100 GPUs over 90-100 days~\cite{walker2023everything}. Given that training happens across many GPUs over an extended period, there is a high likelihood that of interruptions because of hardware failures, software bugs, or capacity management issues. For instance, the 54-day training of LLaMA 3.1~\cite{llama3TechReport} on 16,000 GPUs encountered 419 failures, with training failing on average every three hours.
In such cases, the model either faces prolonged and unpredictable failure recovery time or must adapt to a new training environment, such as continuing training with a different number of GPUs or migrating from one training cluster to another. However, existing parallelism strategies are hard to adapt, which calls for innovations that enable \emph{reconfigurable parallelism}, which allows a training job to flexibly reconfigure its parallelism strategies while maintaining the same training convergence and accuracy.  

One promising solution to enable reconfigurable parallelism is through the checkpointing system. Intuitively, a model checkpoint contains a snapshot of the model state, including weights, optimizer states, and additional metadata that captures the training progress (e.g., training iterations). If a checkpoint can be saved from one type of parallelism strategy (e.g., ZeRO-style data parallelism~\cite{zero-optimizer}) and reloaded to resume training with a different strategy (e.g., 3D parallelism~\cite{megatron-lm-v2}), it becomes possible to reconfigure parallelism across different hardware configurations. \revise{This results in more resilient DNN training in the face of hardware failures and allows for more flexible parallelism choices to adapt to resource elasticity.} Popular DL frameworks, such as PyTorch~\cite{dcp}, Megatron~\cite{megatron-distributed-checkpointing}, and DeepSpeed~\cite{deepspeed}, already have their checkpointing systems and some research work has also proposed efficient methods for checkpointing~\cite{check-n-run,checkfreq,gemini}. However, are they sufficient to provide such reconfigurable parallelism?

\paragraph{Challenges of reconfigurable parallelism.} Unfortunately, existing checkpointing systems are limited in supporting reconfigurable parallelism for three main reasons.
First, existing checkpointing mechanisms are highly coupled to specific parallelism strategies~\cite{checkfreq,gemini,dcp,mcp}. As a result, different parallelism strategies create distributed checkpoint files in different structures, and a checkpoint for a specific parallelism strategy cannot be loaded for another strategy.
For instance, PyTorch Distributed Elastic~\cite{pytorch-elastic} only allows adjustment of data parallel degrees (i.e., the number of data parallel workers), and there is currently no system that supports reconfiguring a checkpoint from ZeRO-style data parallelism~\cite{zero-optimizer} to 3D parallelism~\cite{megatron-lm-v2} or reconfiguring hybrid parallelism for sparse Mixture-of-Experts models~\cite{mixtral}.

Second, there lacks an automatic reconfiguration pipeline for a wide range of commonly used parallelism strategies.
Because existing checkpointing systems are tightly coupled with specific parallelism strategies, only a small subset of these strategies can be reconfigured through checkpoints. This is typically achieved by implementing dedicated converters~\cite{dcp,mcp} or ad-hoc conversion scripts~\cite{mamba-transformer-script}. However, this approach requires significant engineering effort and system expertise while still being error-prone, because it requires navigating a large and complex set of parallelism strategies and carefully examining how each strategy maps to model checkpoints. This includes determining whether each tensor operator in a model has its data replicated or partitioned, identifying the axis along which an operator is partitioned, and understanding how these replicated parameters and partitioned parameters map to distributed checkpoint files.

Third, as model sizes continue to increase, the overhead associated with reconfiguring from one parallelism to another becomes non-negligible. Larger models often require more time for reconfiguration, which can lead to sub-optimal training performance if not optimized. Therefore, it is important to ensure that the reconfiguration process is efficient and scalable. While there has been work optimizing the checkpoint saving and loading overhead~\cite{checkfreq,check-n-run,gemini,fast-persist}, few studies have focused on optimizing the reconfiguration overhead of distributed checkpoints.


\paragraph{Our approach.} To address the aforementioned challenges, we take a different approach by introducing \emph{Universal Checkpointing} (UCP), a novel checkpointing system that enables large-scale DL training with reconfigurable parallelism. 
UCP allows users to flexibly reconfigure a significantly larger and more complicated set of parallelism strategies compared to prior checkpointing mechanisms, including widely used parallelism strategies such as DP~\cite{ddp}, TP~\cite{megatron-lm}, PP~\cite{gpipe}, SP~\cite{sp,deepspeed-ulysses}, ZeRO~\cite{zero-optimizer}, and their combinations such as 3D-Parallelism~\cite{megatron-lm-v2,mt530b} with variable GPU counts. \UCP achieves this through several techniques. First, we introduce \emph{atom checkpoint}, which is a new checkpoint structure that decouples the checkpoint file from any specific parallelism strategy and hardware configurations, while being flexible to adapt to a wide range of parallelism strategies. Second, we introduce a \emph{pattern-based reconfiguration pipeline} that enables reconfigurable parallelism on top of atomic checkpoints via a well designed \emph{pattern set} and \emph{pattern-based reconfiguration operations}. 
We show that this pipeline allows automatically and flexibly reconfiguring complex parallelism strategies such as 3D parallelism and ZeRO-style data parallelism over state-of-the-art LLM architectures, including dense LLMs, and architectures that require hybrid parallelisms, such as sparse Mixture-of-Experts (MoE)~\cite{mistral-7b}, and models with irregular attention such as Grouped-Query Attention (GQA)~\cite{llama3}. Finally, we introduce efficient system optimizations to reduce the reconfiguration cost via \emph{nested parallel reconfiguration}, \emph{redundancy-bypassing loading}, and \emph{lazy reconfiguration invocation}, which makes impact of \UCP on the training negligible. 

We conduct extensive evaluation on large-scale LLM models, including both dense GPT~\cite{megatron-lm} models and sparse MoEs~\cite{mistral-7b}.
Our evaluation results show that \UCP enables reconfiguration of a significantly larger and more complex set of parallelism strategies from arbitrary distributed checkpoint files, outperforming prior checkpointing mechanisms. Additionally, we show that \UCP is efficient and scalable, which adds minimal reconfiguration overhead. To summarize, we make the following contributions:
\begin{itemize}
    \item We formulate the problem of reconfigurable parallelism for large-scale distributed training and develop Universal Checkpointing (UCP), a novel checkpointing system that enables reconfigurable parallelism for a broad set of commonly used parallelism strategies.
    \item We introduce atomic checkpoint, a new checkpoint structure that decouples checkpoints from specific parallelism strategies and serves as a common representation for flexible reconfigurable parallelism.
    \item We design a pattern-based reconfiguration pipeline, which provides systematic and automated parallelism reconfiguration through a carefully designed pattern set and pattern-based reconfiguration operations. 
    \item We introduce nested parallel reconfiguration and lazy reconfiguration invocation, which significantly reduce the reconfiguration cost. Compared to the sequential approach used in ad-hoc conversion scripts, our nested-parallel method achieves a 14-257x reduction in time cost for models ranging from 7B to 1T parameters. The reconfiguration cost for 1T model is within 3 minutes, which is less than 0.001\% of the total training time.
    \item We conduct extensive evaluation on models with multi-billion parameters and demonstrate the effectiveness of \UCP in terms of accuracy, coverage, and efficiency. 
\end{itemize}
\UCP has been verified through the end-to-end training of several real-world large-scale models, \revise{including BigScience BLOOM (176B)~\cite{bloom}, Microsoft Phi-3.5-MoE (42B)~\cite{phi3}, UCB SmileyLlama (8B)~\cite{smileyllama} and RUC YuLan-Mini (4.2B)~\cite{yulan}}.
This has greatly improved these models' resilience to hardware failures during training, reducing their failure waiting time. It also provides these models flexibility to choose a better-suited parallelism strategy when the training environment changes via reconfigurable parallelism. We have implemented and open-sourced \UCP in \revise{DeepSpeed\footnotemark}. \UCP also becomes readily usable through popular DL frameworks that integrate the library as a backend, such as HuggingFace~\cite{huggingface} and PyTorch lightning~\cite{pytorch-lightning}.

\footnotetext[1]{\href{https://github.com/deepspeedai/DeepSpeed}{https://github.com/deepspeedai/DeepSpeed}}

\section{Background}
\label{sec:background}

We provide a brief introduction to the background of parallelism strategies  for distributed training. In particular, training DNNs efficiently at scale typically involves the following parallelism strategies.

\paragraph{Data Parallelism (DP).} DP is one of the most commonly used parallelism strategies for scaling DNN training on multi-GPUs. It divides a mini-batch into micro-batches across data parallel workers but the model state is replicated. Each worker independently computes a micro-batch and calculates gradients, which are subsequently synchronized through all-reduce to get consistent model parameters throughout training. To eliminate the redundancy of replicated model states in DP, zero redundancy optimizer (ZeRO) was introduced to progressively eliminate model redundancy in optimizer states (ZeRO-1), gradients (ZeRO-2), and weights (ZeRO-3)~\cite{zero-optimizer}. 

\paragraph{Tensor Parallelism (TP).} To train models that exceed device memory, TP partitions compute and memory-intensive tensor operations (e.g., \emph{matmul}) along non-batch axes across multiple devices. It then lets each GPU compute the partitioned operation in parallel and uses communication collectives (e.g., \emph{all-reduce}, \emph{all-gather}) at the split and merge point of these partitioned operators to ensure consistent states~\cite{megatron-lm}.  

\paragraph{Pipeline Parallelism (PP).} Different from TP, PP divides model states into multiple stages on different GPUs~\cite{gpipe,1f1b}. Meanwhile, a pipeline scheduler splits a mini-batch into micro-batches and streams the micro-batches through different pipeline stages. Two consecutive pipeline stages exchange intermediate data via point-to-point communication. Such data dependencies lead to pipeline bubbles, i.e, computation stalls. Multiple pipeline schedules such as 1F1B~\cite{1f1b} and interleaved 1F1B~\cite{megatron-lm-v2} have been introduced to reduce pipeline bubbles to improve overall pipeline efficiency. 

\paragraph{Sequence Parallelism (SP).} Motivated by training long-context LLMs, researchers have also developed SP, which divides the model activations along the sequence input dimension and leverages techniques such as distributed attention mechanism to hold the large activation memory from training long sequence inputs~\cite{sp,deepspeed-ulysses,ring-attention}.

These parallelism strategies are often combined to scale out modern DNN training, and a training system needs to orchestrate different parallelism strategies with the goal to achieve high training efficiency. For example, the state-of-the-art training systems such as Megatron-LM and DeepSpeed, combine these parallelism strategies for scaling Transformer-based LLMs, known as 3D/4D parallelism~\cite{megatron-lm-v2,mt530b,megascale}.

\section{Problem Formulation}
\label{sec:problem}

In this section, we formulate the problem of reconfigurable parallelism for distributed DL training. In particular, as input, we are given (1) a training task that has a DNN model $\Phi$ with a model state of $\Phi(t)$ (weights and optimizer states) at training iteration $t$, (2) a \emph{Source} parallelism strategy $P_{src}$ that decides how $\Phi(t)$ is partitioned and executed in parallel on $N_{src}$ parallel workers (e.g., 64 NVidia A100 GPUs), 
(3) a \emph{Target} parallelism strategy, on a target hardware configuration with $N_{tgt}$ workers (e.g., 48 NVidia A100 GPUs), to which the model training needs to be reconfigured. 
In both cases, a parallelism strategy can be DP~\cite{ddp}, TP~\cite{megatron-lm}, PP~\cite{gpipe,1f1b}, SP~\cite{sp}, ZeRO~\cite{zero-optimizer}, or a composition of those strategies such as 3D/4D parallelism~\cite{megatron-lm-v2,mt530b}. 
The objective is to design a reconfiguration mechanism that transforms $\Phi(t)$ from the \emph{Source} parallelism strategy $P_{src}$ to the target strategy $P_{tgt}$.

\section{Challenges and Opportunities}
\label{sec:motivation}

In this section, we introduce the challenges prior work faces to offer reconfigurable parallelism and opportunities for improvement, which guides the design in the subsequent section. 

\begin{figure}[!t]
    \centering
    \includegraphics[width=0.9\linewidth]{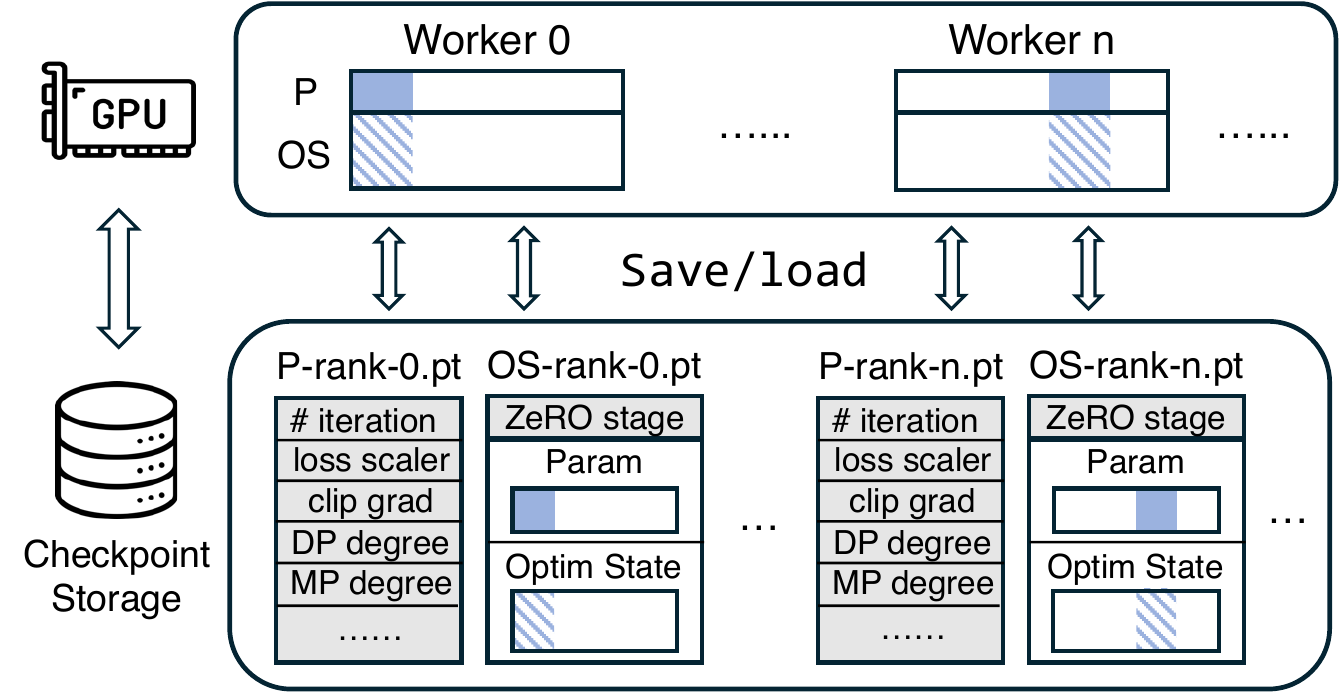}
    \caption{Existing distributed checkpoints are highly coupled to specific model parallelism strategy. Each worker in ZeRO-3 saves its own unique sharded model parameters (P) and optimizer states (OS). Reconfiguring ZeRO-3 requires altering each sharded parameter and optimizer states.}
    \label{fig:coupled-chpt}
\end{figure}

\paragraph{Model parallelism implementation-coupled checkpoint saving/loading.}
The distributed checkpoints are highly coupled to specific model parallelism strategy. For example, data parallelism (e.g., PyTorch DistributedDataParallel~\cite{ddp}), one of the most commonly used parallelism strategies, replicates model weights and optimizer states across GPUs, and therefore only one rank (e.g., rank 0) saves the entire model states in a checkpoint file. When resuming training, each data parallel worker loads the same checkpoint file before launching the training iterations. Other parallelism techniques, such as 
ZeRO~\cite{zero-optimizer} and 
3D parallelism~\cite{megatron-lm-v2},
shard model parameters and optimizer states across GPUs. Since the model states are partitioned across GPUs, each GPU separately saves and loads checkpoint files that contain only a fraction of model state it owns. 
\fref{fig:coupled-chpt} shows the structure of checkpoint files from ZeRO-3, which is very different from others.

Model parallelism-coupled distributed checkpoints lead to a relatively simple checkpointing saving/loading pipeline, because the checkpointing loading is simply the reverse of checkpoint saving process for each worker. 
This design also carries a performance advantage because distributed checkpoint saving requires to block all parallel workers, and having each parallel worker saves the model states it owns incurs no additional synchronization overhead and does not negatively impact the overall training speed. 
However, this coupled design makes it painful and error-prone to reconfigure the checkpoints to a different parallelism strategy, because developers need to hand-write conversion scripts to convert the checkpoint files from one parallelism to another. As a result, existing frameworks support reconfiguration of only a very limited set of parallelism strategies, e.g., changing the data parallel degree in distributed data parallel~\cite{pytorch-distributed-checkpointing}, 
or supports weight-only conversion of distributed checkpoints for inference-only evaluation (e.g., the conversion of optimizer states is not supported)~\cite{megatron-distributed-checkpointing}.
As such, any node failure or resource reallocation in the training process will hobble the entire training process if there lacks support for a training job to reconfigure its parallelism on a new hardware environment. 


\begin{figure*}[!th]
\centering
\includegraphics[width=0.86\linewidth]{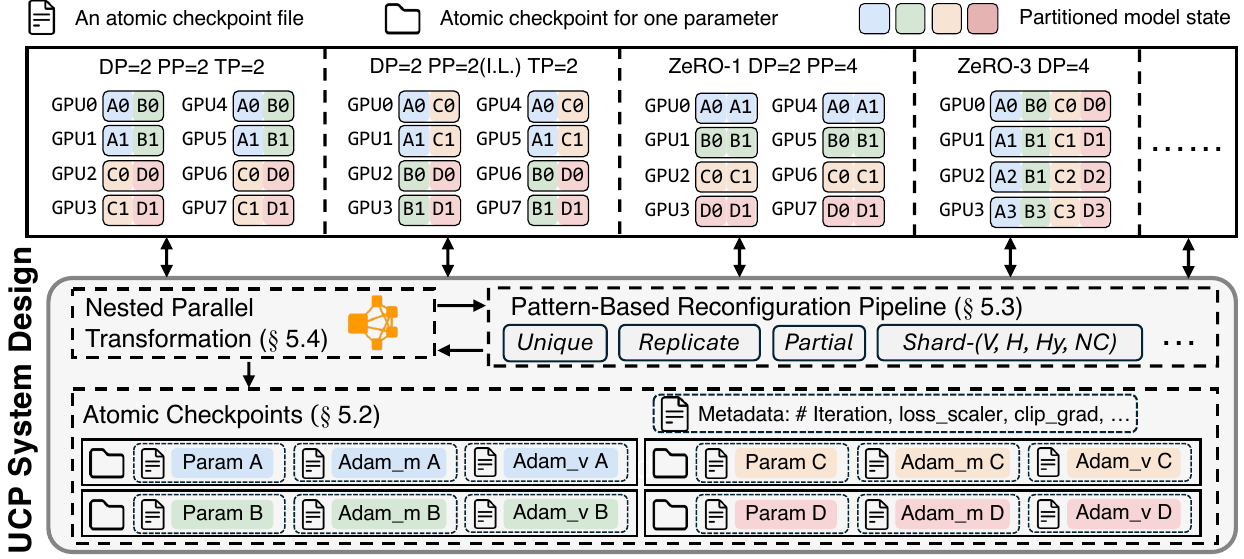}
\caption{Overview of \UCP system design. \UCP enables flexible and efficient reconfiguration from any \emph{Source} parallelism strategy, e.g., $P_{src}$ (ZeRO-1 + PP) = (DP=2, PP=4) to any \emph{Target} parallelism strategy, e.g., $P_{tgt}$ (3D-Parallel) = (DP=2, TP=2, PP=2), via {atomic checkpoints}, {pattern-based reconfiguration pipeline}, and efficient optimizations such as {nested parallel reconfiguration}.} 
\label{fig:overview}
\end{figure*}

\paragraph{Limited coverage of automatic reconfiguration pipeline for a large set of widely used parallelism strategies.} 
Table~\ref{tab:comparison} summarizes the state-of-the-art checkpointing systems for DL training.
\revise{CheckFreq~\cite{checkfreq} is an efficient checkpointing system that overlaps the computation of training with checkpoint saving and adaptively tunes the checkpointing frequency at runtime based on I/O profiling.} Gemini~\cite{gemini} introduces an in-memory checkpointing technique that checkpoints GPU state to local and remote CPUs and interleaves checkpointing IO with training computation to reduce the overhead, which impressively enables checkpointing on every training iteration. Both CheckFreq and Gemini accelerate checkpointing efficiency but do not provide support for reconfiguration of advanced model parallelism such as TP, PP, and ZeRO. Some concurrent work have proposed changing parallelism through checkpointing. 
For example, PyTorch Distributed Checkpointing (DCP)
supports changing the parallelism degrees via loading-time resharding, e.g., after all distributed checkpoint files are loaded to the CPU, each sharded parameter with a shape of $[X/n, Y]$ is concatenated to obtain the complete $[X, Y]$ parameter. The tensor is then resharded based on the current DP degree.
Similarly, Megatron Dist Checkpointing (MCP) saves checkpoints with a local shape of $[PP*TP, X/PP, Y/TP]$. During loading, these checkpoints are concatenated to a global shape of $[1, X, Y]$ and then sharded based on the current PP and TP degrees.
Both of these approaches are limited in that they only support a subset of parallelism strategies. Given that different parallelism strategies have different performance implications on various hardware, it is desirable to have reconfigurable parallelism that covers widely used parallelism strategies
that are also easily extensible to different model architectures.

\begin{table}[t]
\centering
\small
\setlength{\tabcolsep}{2.8pt}
\begin{tabular}{lccccc}
\toprule
 & DP~\cite{ddp} & TP~\cite{megatron-lm-v2} & PP~\cite{1f1b} & SP~\cite{sp} & ZeRO~\cite{zero-optimizer} \\
\midrule
CheckFreq~\cite{checkfreq} & \xmark & \xmark & \xmark & \xmark & \xmark \\
Gemini~\cite{gemini} & \cmark & \xmark & \xmark & \xmark & \xmark \\
DCP\footnotemark & \cmark & \xmark & \xmark & \xmark & \cmark \\
MCP\footnotemark & \cmark & \pmark & \cmark & \xmark & \xmark  \\
UCP & \cmark & \cmark & \cmark & \cmark & \cmark \\
\bottomrule
\end{tabular}
\caption{Comparison of support for reconfiguring distributed checkpoints across various parallelism strategies. Triangle(\pmark) represents partial support with limitations.}
\label{tab:comparison}
\end{table}

\footnotetext[2]{\href{https://pytorch.org/tutorials/recipes/distributed\_checkpoint\_recipe.html}{https://pytorch.org/tutorials/recipes/distributed\_checkpoint\_recipe.html}}
\footnotetext[3]{\href{https://docs.nvidia.com/megatron-core/developer-guide/latest/api-guide/dist\_checkpointing.html}{https://docs.nvidia.com/megatron-core/developer-guide/latest/api-guide/dist\_checkpointing.html}}

\paragraph{High reconfiguration overhead.} Modern DNN sizes are often massive due to their large number of parameters, e.g., Transformers~\cite{transformer,scaleing-law-nlp} have increased by over 1000$\times$ in the past few years. 
Notable examples include PaLM~\cite{pathway} with 540 billion parameters and MT-NLG~\cite{mt530b} with 530 billion parameters. The checkpoint size is approximately 12$\times$ the parameter count (4$\times$ for model parameters and 8$\times$ for optimizer states), thus checkpoint for these models reaches tens of TB in size. 
Given the huge checkpoint sizes, the reconfiguration time can be constrained by the bandwidth of the persistent storage~\cite{check-n-run,gemini}. For example, it takes 42 minutes to checkpoint the model states of MT-NLG~\cite{mt530b} to the remote persistent storage when the bandwidth is 20Gbps. Given the large and ever increasing model sizes, reconfiguration needs to have relatively low overhead in order to avoid becoming a major performance bottleneck. 

\section{Universal Checkpointing}

\subsection{UCP Design Overview}

\UCP is a distributed checkpointing system, specifically designed for handling large-scale DL training with reconfigurable parallelism. It allows flexible and efficient reconfiguration from any \emph{Source} parallelism strategy to any \emph{Target} parallelism strategy. Since training accuracy is the main requirement, we focus on design choices that do not lead to training accuracy loss or related metric degradation. We first provide an overview of \UCP and discuss each component in more details in subsequent sections. 

\fref{fig:overview} provides an overview of \UCP. At a high level, \UCP consists of three major components: atomic checkpoints, pattern-based reconfiguration pipeline, and nested parallel transformation. In \sref{sec:motivation}, we describe the challenge of parallelism-coupled checkpoints, which are cumbersome to reconfigure. We address this challenge by raising the level of abstraction, and decoupling the checkpoint file format from its parallelism strategy and hardware configurations through \emph{atomic checkpoints} (\sref{subsec:ucp_format}). Atomic checkpoints enable radically simpler mechanism to achieve flexible reconfiguration, spanning the space of parallelism choices. \UCP also provides a pattern-based reconfiguration pipeline to systematically and automatically reconfigure distributed checkpoints from a \emph{Source} parallelism to a \emph{Target} parallelism through atomic checkpoints (\sref{sec:design:ucppipe}). \UCP achieves this by carefully designing a parallelism pattern set and developing pattern-aware reconfiguration operations to automatically convert \emph{Source} distributed checkpoints into atomic checkpoints and map atomic checkpoints to any \emph{Target} parallelism strategy. \UCP leverages \emph{nested parallel reconfiguration}, \emph{redundancy-bypassing loading} and \emph{lazy reconfiguration invocation} to reduce the reconfiguration overhead such that reconfiguration only adds minimal overhead to the training process (\sref{subsec:efficient-impl}). 

\subsection{Atomic Checkpoint}
\label{subsec:ucp_format}


In existing DL distributed training systems, different parallelism strategy handles model states differently. For example, DP replicates and maintains a consistent view of model states during training across data parallel workers, so it is enough to have a single GPU for checkpointing. Other parallelism strategies, such as tensor-slicing parallelism or pipeline parallelism, partition model states along different axes and layers across GPUs. Therefore, each GPU creates a checkpoint file based on a snapshot of the partitioned model states that are in its local memory. Because of this highly parallelism-coupled checkpointing mechanism, practitioners need to hand-write conversion scripts to convert distributed checkpoints for different parallelism strategies, which incurs significant engineering cost and hard to maintain.

In \UCP, we use \emph{atomic checkpoints} to represent model states. An \emph{atomic checkpoint} contains a consolidated view of the model states corresponding to a tensor operation, e.g., a merged state of fragmented states, if the tensor operation is partitioned over multiple parallel workers. Different model states (e.g., weights and optimizer states) have separate atomic checkpoint files, and all atomic checkpoint files corresponding to one parameter is considered as a conjunction of atomic checkpoints (e.g., a directory) to that parameter. 
\revise{In line with standard practices in mixed-precision training, UCP stores optimizer states in FP32 to maintain numerical stability and avoid loss of precision. While this increases checkpoint size, it ensures correctness during training.}
Without loss of generality, assuming the training uses the Adam optimizer~\cite{adam}, for each tensor operation, a conjunction of atomic checkpoints to its parameter has three atomic checkpoint files:
\begin{description}
    \item[model.pt:] A fp32 tensor represents parameter's weight.
    \item[adam\_m.pt:] A fp32 tensor represents first order moment.
    \item[adam\_v.pt:] A fp32 tensor represents second order moment.
\end{description}

The atomic checkpoint is much more fine-grained than most existing distributed checkpoints that have the entire local snapshot of individual parallel worker. An atomic checkpoint is also no longer coupled with any parallelism strategy or hardware configurations, e.g., it does not include any rank id, partitioning information, or additional data from specific sharding strategies such as padding for alignment. It simply maps each parameter in the model to a consolidated view of its weights and corresponding optimizer states. 

The atomic checkpoint is sufficient to map to a wide range of parallelism algorithms, and it also enables flexible reconfiguration of parallelism through checkpoints. Critically,
this representation is naturally data parallel when each data parallel worker needs to load a replica of the model states in a parameter-by-parameter basis. Also, as we identify partitioning patterns in many parallelism (\sref{subsub:pattern}), it becomes possible to define a set of primitives to automatically transform atomic checkpoints into specific parallelism strategies (\sref{subsubsec:pattern-based}). 
As such, \UCP acts as a common interchange format between different distributed training techniques, and one does not need to implement individual converters from each \emph{Source} to \emph{Target} parallel config, significantly increasing the flexibility and coverage of reconfigurable parallelism. 

\hide{Parallel Processing: The format supports parallel data transfer and computation, crucial for minimizing I/O latency and maximizing throughput.
Flexibility: The granular nature of atoms allows for more adaptable checkpoint management, enabling partial updates and efficient access to specific model components.
Efficiency: By aligning with modern hardware capabilities, such as multi-core processors and high-bandwidth storage interfaces, this format optimizes resource utilization.
Scalability: As model sizes continue to grow, this approach scales more effectively than monolithic storage solutions, accommodating the increasing demands of large-scale AI systems.}

\paragraph{Special consideration for data types.} In practice, developers often adopt hardware-friendly training techniques such as mixed-precision training (MPT)~\cite{mixed-precision-training}, where both IEEE float16 weights and float32 weights and optimizer states are maintained to leverage the high computation throughput of hardware features such as TensorCore~\cite{tensorcore}. 
MPT affects the design of reconfigurable parallelism, as different states may have different data formats. Furthermore, dynamic mixed precision training saves checkpoints in one data format (e.g., float16) and switches to another one (e.g., bfloat16 or float32) at different training stages, which have different bit allocation policies between the mantissa and exponent~\cite{bfloat16}.
Therefore, it is crucial to consider the data formats for checkpoints to accommodate these training techniques.  In \UCP, we keep all atomic checkpoints (weight/optimizer values) in float32, and add support to resume training in other data types, including float32, float16, and bfloat16, making it flexible to reconfigure parallelism with different data types.


\vspace{10pt}
\subsection{Pattern-Based Reconfiguration \Pipe}
\label{sec:design:ucppipe}

Atomic checkpoints are decoupled from specific parallelism strategies. However, another important design choice remains unspecified: How to automatically convert distributed checkpoints \emph{from} various $P_{src}$ to atomic checkpoints and map atomic checkpoints to different $P_{tgt}$ parallelism strategies? Most importantly, how to make sure that this automatic conversion process covers a wide range of widely used parallelism strategies as well as new parallelism strategies from state-of-the-art model architectures? 
The design choice here does not change the meaning of atomic checkpoints, but they are essential to ensure the utility of the resulting reconfiguration system. To provide answers, in this section, we develop a \emph{pattern-based reconfiguration pipeline}.

\vspace{5pt}
\subsubsection{Designing the Pattern Set} 
\label{subsub:pattern}
To support flexible reconfiguration of parallelism based on atomic checkpoints, we seek an approach that can automatically map distributed checkpoints into atomic checkpoints. To achieve that, we look into patterns inside distributed checkpoint files so we can perform pattern-based transformation. The key consideration here is how to design and select the patterns. Good patterns should be able to have two key properties: coverage and flexibility. The coverage is desirable because it enables mapping between distributed checkpoints from a wide range of parallelism strategies and model architectures to atomic checkpoints. The flexibility facilitates subsequent parallelism reconfiguration operations. Given these two key properties, we propose the following pattern set.  

\begin{figure}[!t]
\centering
\includegraphics[width=0.9\linewidth]{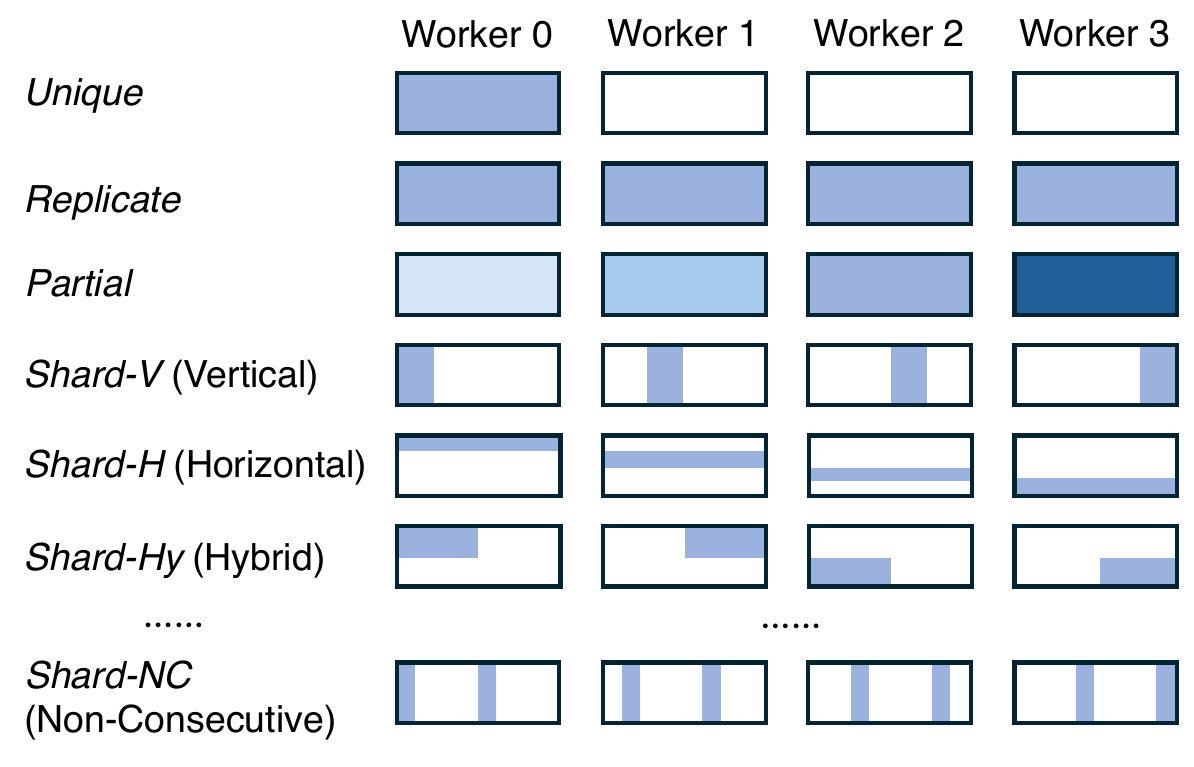}
\caption{Illustration of patterns defined in \UCP.}
\label{fig:pattern_complex}
\end{figure}

\fref{fig:pattern_complex} illustrates the pre-defined patterns in \UCP. \uniquepattern means that a tensor is uniquely associated with a distributed checkpoint, which is commonly seen from checkpoints generated through inter-op parallelism such as \pp.
\replicatepattern means one parameter will be replicated across multiple distributed checkpoints. For example, even though \emph{matmuls} are sharded across GPUs in TP, other parameters such as LayerNorm and the Bias terms are replicated across GPUs because they are not compute and memory intensive. As a result, those replicated parameters will cause redundancy if we converted all those replicated parameters to atomic checkpoints.  
\partialpattern indicates a parameter that is updated independently across GPUs, which corresponds to models with asynchronous training. 
More complicated patterns exist in intra-op parallelism strategies, which \revise{partition} parameters along certain dimensions, e.g., \emph{Shard-V} for column-wise sharding and \emph{Shard-H} for row-wise sharding in \tp, \emph{Shard-Hy}, which shards along multiple dimensions of the model states, and \emph{Shard-NC}, which shards model states non-consecutively. We show later that the set of patterns we choose enables conversion of complicated parallelism strategies such as ZeRO-3~\cite{zero-optimizer} and 3D parallelism~\cite{megatron-lm-v2} while being flexible enough to also support new model architectures such as MoE~\cite{mixtral} and GQA~\cite{gqa}.
Note that although the set of patterns covers many existing parallelism strategies, this approach is general and extensible to support new parallelism patterns, not just the ones considered in the paper.


\vspace{5pt}
\subsubsection{Pattern-Aware Reconfiguration Operations}
\label{subsubsec:pattern-based}
Given the pattern set, \revise{\UCP extracts} the pattern information from distributed checkpoints and launches pattern-aware reconfiguration operations. 
In particular, Table~\ref{tab:ucp_primitives} lists the main reconfiguration operations in \UCP, including \extract, \union, \padfree, \genmeta, and \load. For different patterns, each operation performs pattern-aware transformation, and Algorithm~\ref{algo:ucp-conversion} demonstrates how the \extract and \union consolidate different parameter fragments based on their patterns.   
We use several concrete examples to illustrate how pattern-aware reconfiguration operations in \UCP support complex parallelism strategies and different model architectures. 

\begin{table}[!h]
\caption{
Pattern-aware reconfiguration operations in \UCP. 
}
\centering
\label{tab:ucp_primitives}
\small
\begin{tabular}{p{0.13\linewidth}|p{0.75\linewidth}}
\toprule
Operator & Definition \\
\midrule
\extract & Takes distributed checkpoint files from $P_{src}$ as input and returns a set of parameter fragments contained in that checkpoint file. \\
\union   & A pattern-specific union is called on parameter fragments to obtain consolidated parameters. \\ 
\padfree   & Strips pattern-specific padding from a consolidated parameter. \\ 
\genmeta   & Generates the metadata (e.g., shape and location information to a given rank) associated with $P_{tgt}$ for each atomic checkpoint.  \\ 
\save   & Saves the consolidated model states for each parameter as atomic checkpoint files.\\
\load   & Loads atomic checkpoints to each rank based on the \genmeta of $P_{tgt}$. \UCP loads atomic checkpoints in a layer-by-layer fashion to prevent model states from exceeding memory limits. \\
\bottomrule

\end{tabular}
\end{table}

\begin{figure}[h]
\centering
\includegraphics[width=\linewidth]{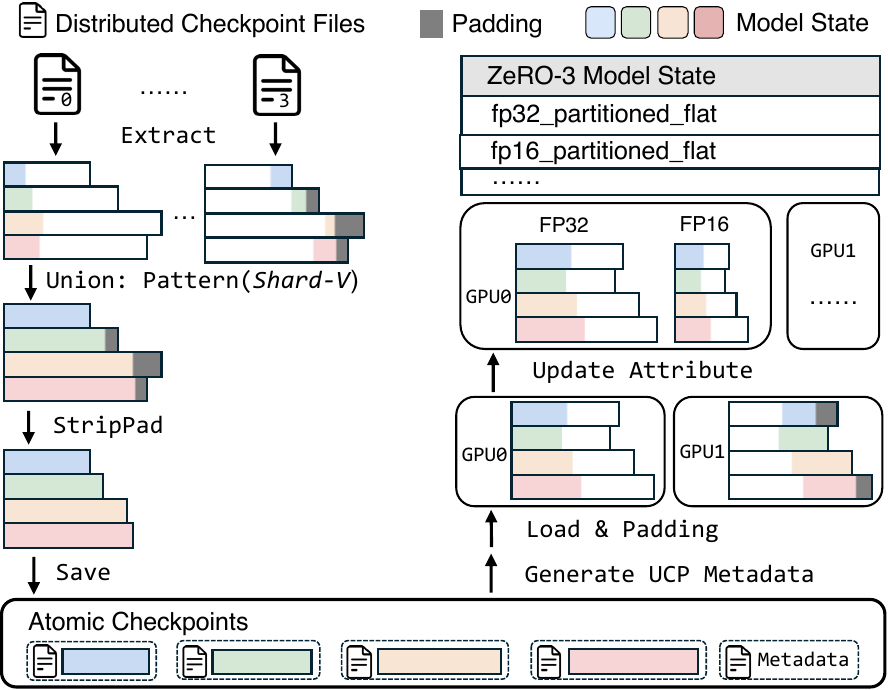}
\caption{Workflow of \UCP to flexibly reconfigure parallelism with ZeRO-3 either as $P_{src}$ or $P_{tgt}$. The left side shows the process of converting checkpoints from $P_{src}$ (ZeRO-3, DP=4) to atomic checkpoints. The right side shows how atomic checkpoints are converted to $P_{tgt}$ (ZeRO-3, DP=2).} 
\label{fig:stage3}
\end{figure}

\paragraph{ZeRO Stage 3.} The process of applying \UCP's reconfiguration pipeline to ZeRO-3 checkpoints is illustrated in \fref{fig:stage3}. If ZeRO-3 is the $P_{src}$, each DP rank persists the sharded parameters and optimizer states it owns to a checkpoint file, leading to $N$ distributed checkpoint files in total when the DP degree is $N$. \UCP then identifies parallelism patterns of ZeRO-3: ZeRO-3 flattens each tensor into a 1D tensor and shards the 1D tensor equally across different data parallel ranks. \UCP identifies those tensors in ZeRO-3 distributed checkpoint files as the \shardpatternv pattern. 
Based on the pattern, \UCP runs \extract and \union on fragmented parameters to create atomic checkpoints, which contain a consolidated view of parameters and optimizer states. 
One complexity of ZeRO-3 is that it adds padding to make sure the 1D tensor can be evenly divided across $N$ DP workers for alignments. For example, for a parameter with a shape [1024], if the original DP degree is 3, then ZeRO-3 pads the parameter to [1026], such that each worker owns a fragment of [342]. 
\UCP uses \padfree to remove the padding and uses \save to save the resulting atomic checkpoints to persistent stage. 

On the other hand, if ZeRO-3 is the $P_{tgt}$, each DP worker calculates its new partition metadata via \genmeta and then loads atomic checkpoints to each rank sequentially, following the layer order and with alignment padding added for high performance. Using the previous example, assuming $N_{tgt} = 2$, \UCP will resume training with each rank having the correct parameter shape [512] instead of [513]. 
Once all the partitioned states are loaded into a GPU, e.g., in the flatten memory attribute \emph{fp32\_partitioned\_groups\_flat} of ZeRO-3, the updated attribute is then broadcast to other necessary attributes, such as \emph{fp16\_partitioned\_groups\_flat} for MPT. Through this process, \UCP flexibly reconfigures parallelism with ZeRO-3 either as $P_{src}$ or $P_{tgt}$ under various DP degrees.

\begin{algorithm}[!t]
\small
\caption{Pattern-Aware Reconfiguration}\label{alg:ucpcp}
\label{algo:ucp-conversion}
\begin{algorithmic}[1]
\Statex $\triangleright$ Extract
\State \quad \textbf{for} distributed checkpoint $ckpt$ in $storage$ \textbf{do in parallel}
\State \quad \quad \textbf{for} param\_name $p$, tensor $t$ in $ckpt$ \textbf{do}
\State \quad \quad \quad \quad \texttt{ToShuffler}($p$, $t$)
\Statex $\triangleright$ Union
\State \quad $param\_list$ = \texttt{FromShuffler()}
\State \quad \textbf{for} param $p$ in $param\_list$ \textbf{do in parallel}
\State \quad \quad  $\{T_1, T_2, ..., T_n\}$ $\leftarrow$ all tensors name matches $p$
\State \quad \quad \textbf{Switch} $p$
\State \quad \quad \quad \textbf{case} \texttt{PatternMatch}(\replicatepattern, $p$) \textbf{then}
\State \quad \quad \quad \quad $ucp_p$ = $T_1$
\State \quad \quad \quad \textbf{case} \texttt{PatternMatch} (\partialpattern, $p$) \textbf{then}
\State \quad \quad \quad \quad $ucp_p$ = \texttt{Sum}($T_1, T_2, ..., T_n$) / $n$
\State \quad \quad \quad \textbf{case} \texttt{PatternMatch}(\shardpattern, $p$) \textbf{then}
\State \quad \quad \quad \quad $ucp_p$ = \texttt{Concat}($T_1, T_2, ..., T_n$)
\State \quad \quad \quad \textbf{case} \texttt{PatternMatch}(\uniquepattern, $p$) \textbf{then}
\State \quad \quad \quad \quad $ucp_p$ = $T$
\State \quad \quad \textbf{if} \emph{hasPadding}($p$) \textbf{then}
\State \quad \quad \quad $ucp_p$ = \texttt{StripPad}($ucp_p$)
\State \quad \quad \texttt{Save}($ucp_p$)
\end{algorithmic}
\end{algorithm}

\paragraph{3D parallelism.} The reconfiguration process with 3D parallelism either as $P_{src}$ or $P_{tgt}$ is similar to ZeRO-3. However, since 3D parallelism consists of various parallelism strategies, it has its own set of complexities. 
Similar to ZeRO-3, when saving distributed checkpoints for 3D parallelism, each GPU only saves a slice of the model state it owns. However, the parameter pattern of 3D parallelism is much more complicated: parameters can have \replicatepattern (e.g., \emph{LayerNorm}), \shardpatternv and \shardpatternh (e.g., \emph{matmul}), \partialpattern (e.g., Alibi Embedding~\cite{alibi}) with TP degree $>$ 1, and \replicatepattern (e.g., Tied-Embedding) pattern and \uniquepattern (majority) with PP degree $>1$. \UCP significantly lifts the burden of manually converting distributed checkpoints from 3D parallelism through pattern matching and using 
\extract, \union, \padfree to create consolidated atomic checkpoints without any padding. For example, depending on \shardpatternv or \shardpatternh, \UCP would concatenate fragmented parameters from TP into a single consolidated tensor either with the row dimension or column dimension. 
\revise{Similar to} ZeRO-3, if 3D parallelism is chosen as the $P_{tgt}$, a new mapping between atomic checkpoints and GPU ranks is generated first through \genmeta, and each rank loads from atomic checkpoints based on the new mapping policy.

Additional complexities from 3D parallelism also come from the pipeline schedules. For instance, Interleaved 1F1B~\cite{megatron-lm-v2} is a pipeline schedule that reduces the pipeline bubbles by assigning non-contiguous layers of a model to the same rank, resulting in distributed checkpoints highly coupled with the specific order of layer assignment.
\UCP's atomic checkpoints are decoupled from such scheduling, as the parameters and optimizer states are only associated with their respective parameter names, allowing to reconfigure the mapping between each rank and the parameters.



\begin{figure}[h]
\centering
        \includegraphics[width=\linewidth]{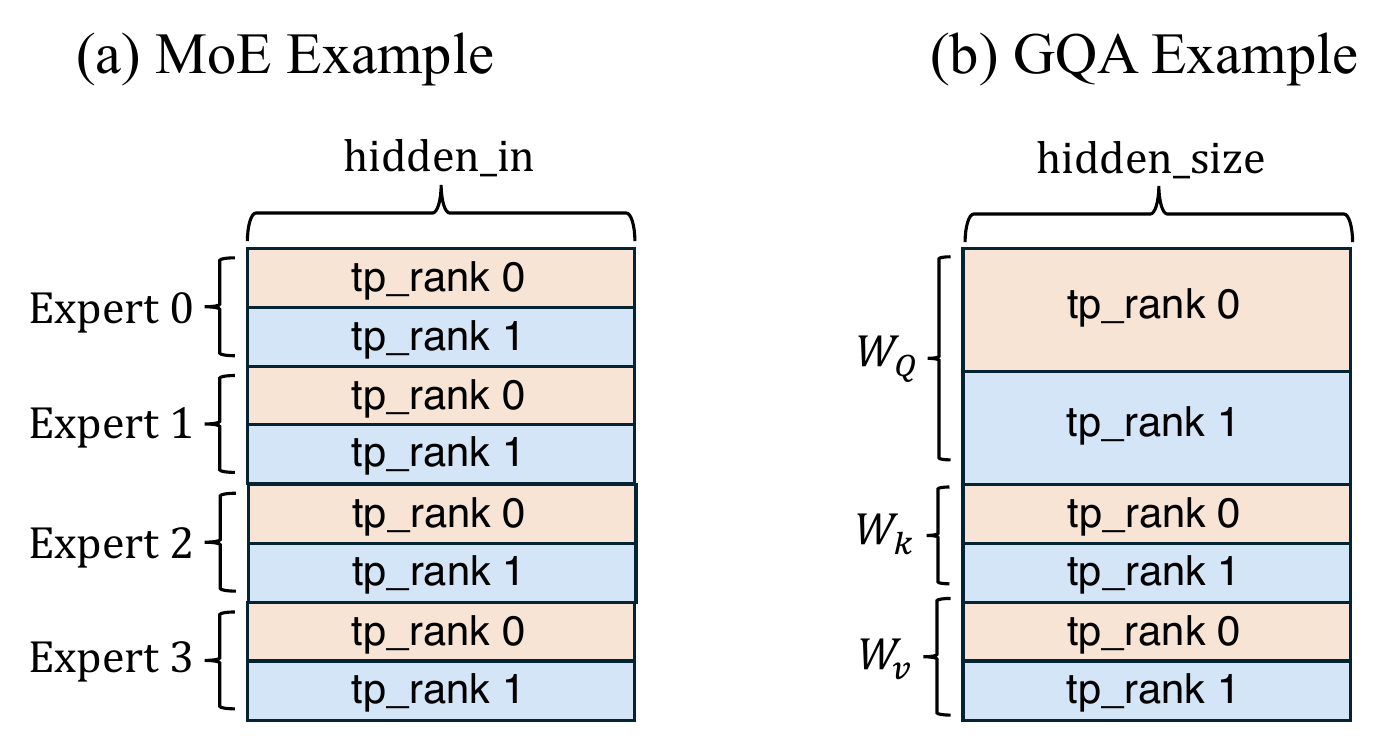}
        \caption{Illustration of the reconfiguration complexities from sparse Mixture-of-Expert models and models with irregular attention mechanisms, such as Grouped Query Attention (GQA). This figure shows examples of MoE with 4 experts and GQA, both with TP=2. The \emph{Shard-NC} pattern allows \UCP to identify partitions along different dimensions and variable-size fragments, and \UCP supports reconfiguration operations based on these identified patterns.}
        \label{fig:sub-pattern}
\end{figure}

\paragraph{Sparse Mixture-of-Experts and Irregular Attention GQA.} The patterns described in \fref{fig:pattern_complex} not only covers model sharding strategies for dense LLMs, as illustrated in the ZeRO-3 and 3D parallelism example, it also covers model architectures such as sparse MoE models~\cite{mixtral} and irregular attention mechanisms such as GQA~\cite{gqa}, both of which have received intensive interests in LLM training but also require more complex parallelism strategies.
Fig.~\ref{fig:sub-pattern} illustrates the complexity. The MoE model in this example defines the weight tensor of an MoE's FFN layer as \emph{[n\_experts $\times$ hidden\_out, hidden\_in]}, which is a \emph{fused} weight matrix different from standard MoEs that use separate matrices to represent different experts. With the fused matrix, one can apply TP to this layer. However, different from standard TP, the partition happens along the \emph{hidden\_out} dimension, which does not fall under the common \shardpatternv and \shardpatternh pattern. In \UCP, this is handled through the \emph{Shard-NC} pattern, which allows \UCP to identify it as a 3-dim tensor and apply pattern-aware reconfiguration operations such as \extract and \union to still obtain a consolidated view of the \emph{fused} matrix from distributed checkpoints. 

In the GQA~\cite{gqa} example, the QKV matrices in the multi-head attention (MHA) are also fused together as one tensor. However, different from traditional MHA, the QKV in GQA have different sizes, causing the fused matrix to have an irregular shape of \emph{[q\_size + k\_size + v\_size, hidden]}, where \emph{q\_size != k\_size == v\_size}. If Tensor Parallelism (TP) is applied to this tensor, it needs to partition the tensor along the first dimension for each Q, K, and V but with different sizes. The \emph{Shard-NC} together with shape info allows \UCP to identify these variable-size fragments and apply pattern-aware reconfiguration operations accordingly. 

These examples show that the pattern-based reconfiguration pipeline in \UCP is quite flexible, supporting various parallelism strategies as either $P_{src}$ or $P_{tgt}$. Meanwhile, its carefully designed pattern set allows it to support not only a wide range of complex parallelism strategies but also very extensible to both sparse Mixture-of-Expert models and models with irregular attention in addition to dense models. 
\revise{Introducing a new parallelism strategy involving patterns beyond UCP’s current scope would require a similar amount of implementation effort as the examples presented above. We actively work with the community to identify these new patterns to extend UCP to have a wider coverage.}

\vspace{20pt}
\subsection{Efficient Reconfiguration}
\label{subsec:efficient-impl}



\UCP is intended for high-performance training, therefore the operators described in \sref{sec:design:ucppipe} need to scale well with larger models and more nodes. To achieve low reconfiguration cost, \UCP leverages three optimizations: (1) Nested parallel reconfiguration, (2) Redundancy-bypassing loading, and (3) Lazy reconfiguration invocation.

\begin{figure}[!t]
    \centering
    \includegraphics[width=0.90\linewidth]{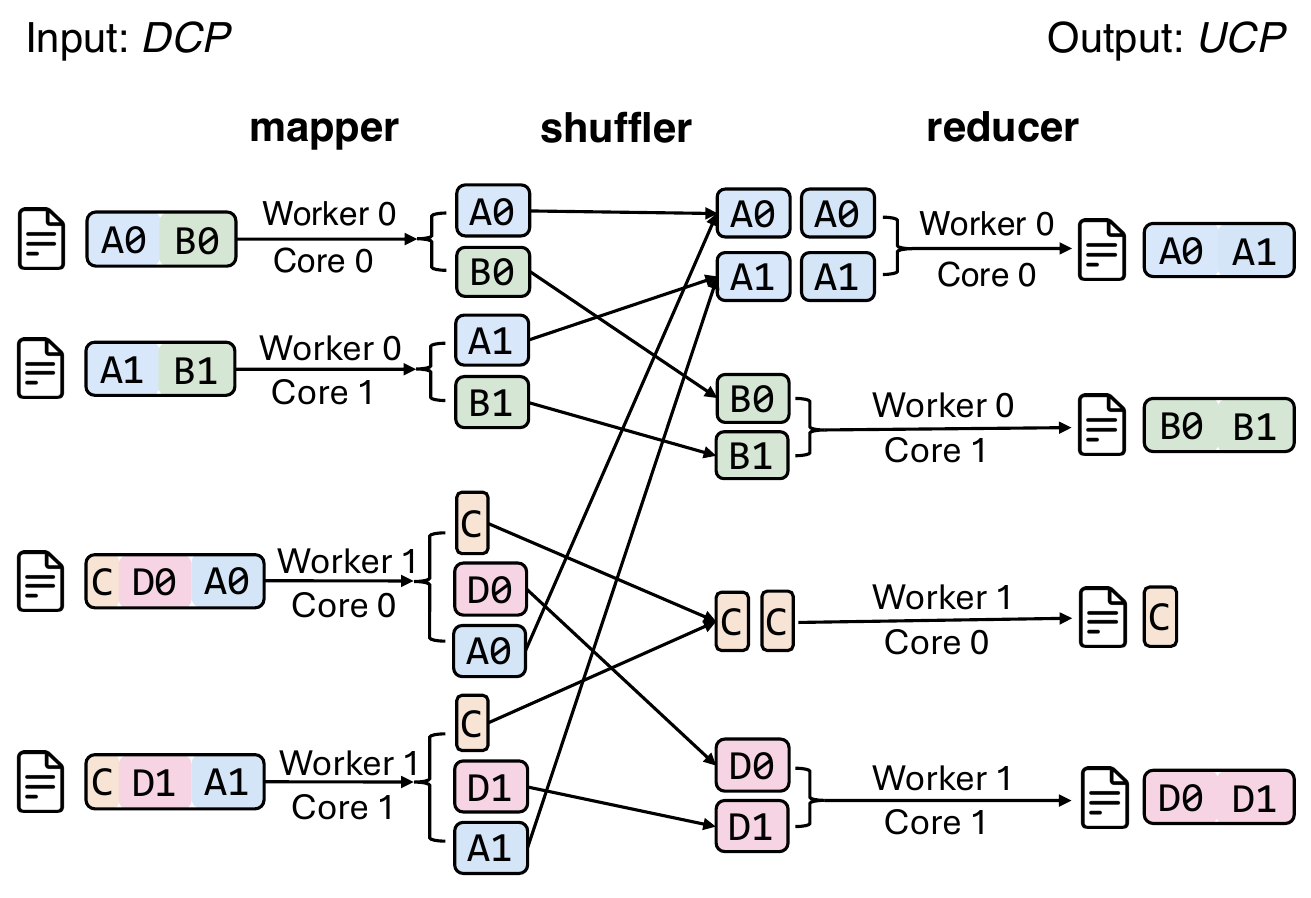}
    \caption{Illustration of the nested parallel reconfiguration process 
    of \UCP. \UCP leverages a MapReduce-based approach to utilize the aggregated compute and bandwidth of multi-node multi-processors to convert distributed checkpoints into atomic checkpoints in parallel. Meanwhile, it performs careful load balancing to avoid the straggle problem.}
    \label{fig:mapreduce}
\end{figure}

\paragraph{Nested Parallel Reconfiguration.} Motivated by leveraging additional nodes to increase the reconfiguration efficiency, we introduce
a nested parallel reconfiguration scheme by formulating the reconfiguration as a MapReduce problem~\cite{mapreduce}. MapReduce was originally introduced as a programming model for processing large datasets in a distributed environment, and has been widely used in big data processing frameworks such as Hadoop~\cite{hadoop}. MapReduce decomposes a problem into three phases: Mapper, Shuffler, and Reducer. In our case, as shown in Fig.~\ref{fig:mapreduce}, each parallel mapper reads a distributed checkpoint and parses it according to the parameter fragments of the tensors it contains.
The shufflers send the output from mappers, i.e., the tensors each representing a fragment of parameter to the corresponding reducers assigned to process those parameters.
The reducers process these parameter fragments based on specific pattern, such as \shardpattern and \partialpattern, which we discussed in detail in \sref{sec:design:ucppipe}.

It is important to note that parameters vary significantly in size. For example, in a 176B model,
the embedding parameter has a size of [50257, 12288], while the bias of LayerNorm has a size of [12288]. Randomly assigning these parameters to workers would lead to imbalance, as both computation time and the loading/saving time are proportional to parameter size. This imbalance results in most workers idly waiting for the one handling the heaviest task, which dominates the overall conversion time. To address this issue, we introduce a pre-calculation based balancing method. The master worker divides parameters into $N$ groups based on \emph{numels}. For instance, with 4 workers, the master might assign the embedding parameter to worker 1, while distributing multiple smaller parameters like LayerNorm biases and weights across workers 2, 3, and 4 to achieve a similar total numel count per worker.

Building upon the balanced distribution of parameters across workers, we have implemented a second level of parallelism utilizing multi-core processing within each worker. This approach leverages the full computational capacity within a worker node by engaging all available CPU cores to process parameters concurrently. The nested parallelism strategy -- balancing across workers and then across cores within each worker -- maximizes resource utilization and minimizes idle time at both the node and core levels, leading to affordable conversion times for large language models.

\paragraph{Redundancy-bypassing loading.}
Each worker within the same DP group shares or \revise{partially shares} (except ZeRO3-DP) the same model states, thus needs to load the same atomic checkpoint files. \UCP eliminates redundant checkpointing loading by evenly distributing loading workloads among workers in the same DP group. Each rank reads its assigned atomic checkpoint files into CPU memory, transfers them to GPU memory, and then uses all-gather operations to distribute the data to other workers. This optimization significantly alleviates the IO bandwidth pressure by eliminating redundant data transfers from storage to CPU memory and leverages the high-bandwidth GPU-to-GPU interconnects (such as NVLINK at 900 GB/s) to redistribute data once it is loaded into GPUs. \revise{Moreover, \UCP employs a memory-efficient loading design. UCP loads atomic checkpoints layer by layer and assigns tensors to GPUs based on the selected parallelism strategy. Once a layer is transferred to GPU memory, its tensors are released from CPU memory, reducing peak CPU memory usage from the size of full model checkpoints to that of a single layer.}


\paragraph{Lazy reconfiguration invocation.} The careful reader may think "when should the reconfiguration of parallelism be triggered?" or "Is it possible to directly consolidate distributed model states into a single checkpoint file when saving checkpoints?" Such a method has two drawbacks: Consolidating distributed model state into a single checkpoint unacceptably slows down training, and is impractical at extreme scales because there may not be sufficient memory to host the consolidated model states. 
To avoid slowing down normal training, we adopt a lazy invocation design for \UCP. 
\UCP reconfiguration is invoked \emph{only} when a $P_{src}$ and $P_{tgt}$ are different or the hardware changes. 
This way, the distributed checkpoint saving logic does not need any change, and \UCP only incurs $<$0.001\% of the training time overhead when there is a need for reconfiguration of parallelism since reconfiguration does not add cost to the critical path length, i.e., the normal distributed training process.

\begin{figure*}[!th]
\centering
\includegraphics[width=\textwidth]{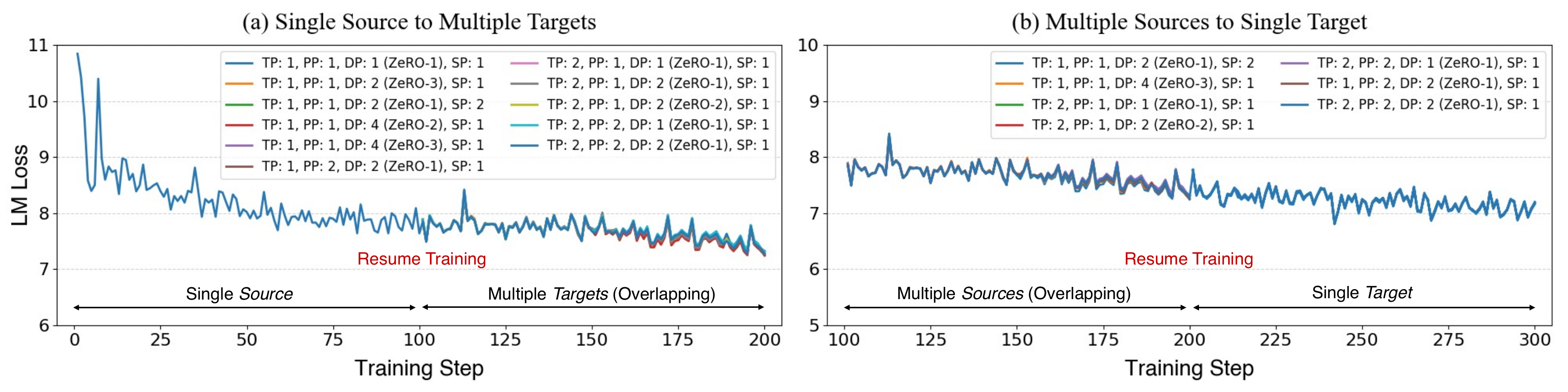}
\vspace{-10pt}
\caption{(a) Training curves of reconfiguring training from one \emph{Source} of parallelism into different \emph{Target} parallelisms. (b) Training curves of reconfiguring training from multiple \emph{Source} parallelism strategies to the same \emph{Target}.}
\vspace{-8pt}
\label{fig:correctness} 
\end{figure*}

\section{Evaluation}
\label{sec:eval}

We implemented \UCP in DeepSpeed and evaluate \UCP through a series of experiments on training LLMs. 
Overall, our evaluation aims to answer the following questions:
\begin{itemize}[leftmargin=*]
    \item Can \UCP enable reconfigurable parallelism without compromising model training accuracy?
    \item How is \UCP compared to existing checkpointing systems in terms of supporting flexible parallelism reconfiguration?
    \item Can \UCP generalize to different model architectures? 
    \item How does \UCP affect the checkpointing saving and reconfiguration overhead?     
\end{itemize}

\subsection{Evaluation Methodology}
\paragraph{Workloads.} For the accuracy evaluation, we focus on evaluating GPT-style Transformer based models. We select several architectures from prior work: GPT-3 medium~\cite{gpt-3} ($L=24, H=1024, A=16, 350M$ params), GPT-3 7B ($L=32, H=4096, A=32, 7B$ params), 
a 176B GPT-3 style model ($L=70, H=14336, A=112, 176B$ params), 
and a Mixtral-7x8B~\cite{mixtral} style MoE model ($L=32, H=3072, A=32, E=16, 42B$ params), to cover different model configurations and model sizes. We use a subset of the Pile dataset~\cite{pile} for training to evaluate the impact to the training loss with and without reconfigured parallelism from \UCP.


\paragraph{Hardware.}
We conducted our experiments on: 64xA100 40GB GPUs (256GB DRAM, 10TB storage, 200Gbps interconnect). The evaluation of the 176B GPT style model was conducted on {384xA100} 80GB GPUs. The efficiency evaluation of the 1TB model was conducted on 1024xMI250X 64GB GPUs. \revise{We measured the I/O bandwidth of the clusters: the NVIDIA GPU clusters achieved 5 GB/s, while the AMD GPU cluster achieved 3 GB/s. These bandwidth levels are comparable to those of consumer-grade SSDs or typical cloud provider offerings.}

\subsection{Evaluation of Reconfigurable Parallelism}

\UCP provides flexible reconfiguration from a \emph{Source} parallelism strategy to a different \emph{Target} varying hardware configurations. 
Since reconfigurable parallelism should not alter the training accuracy (model states preserved after the reconfiguration), we focus first to determine the accuracy impact of \UCP with reconfigurable parallelism. After all, regardless how much reconfigurable parallelism a system offers, if it does not preserve training accuracy, it severely limits its applicability in practice. We conduct the evaluation through two categories of reconfiguration: 

\vspace{-2pt}
\paragraph{Single \emph{Source} to multiple \emph{Target}.} To test if \UCP allows resuming training with different \emph{Target} parallelism strategies and hardware configurations, we first train the GPT-3 model using a configuration of TP=2, PP=2, DP=2 (ZeRO-1), and SP=1. Due to constraints in time and resources, we limited the experiment to the first 200 iterations. We enable \UCP on the checkpoints saved at the 100th iteration and resume training using different parallelism strategies and GPUs. We record the LM loss for each iteration. Fig.~\ref{fig:correctness}(a) illustrates that the training can be seamlessly resumed using different \emph{Target} parallelism strategies, while achieving consistent convergence if the training were to continue with the \emph{Source} strategy.
These results confirm that \UCP enables resuming training to different hardware and parallelism configurations. 

\begin{figure*}[!ht]
\centering
\includegraphics[width=\textwidth]{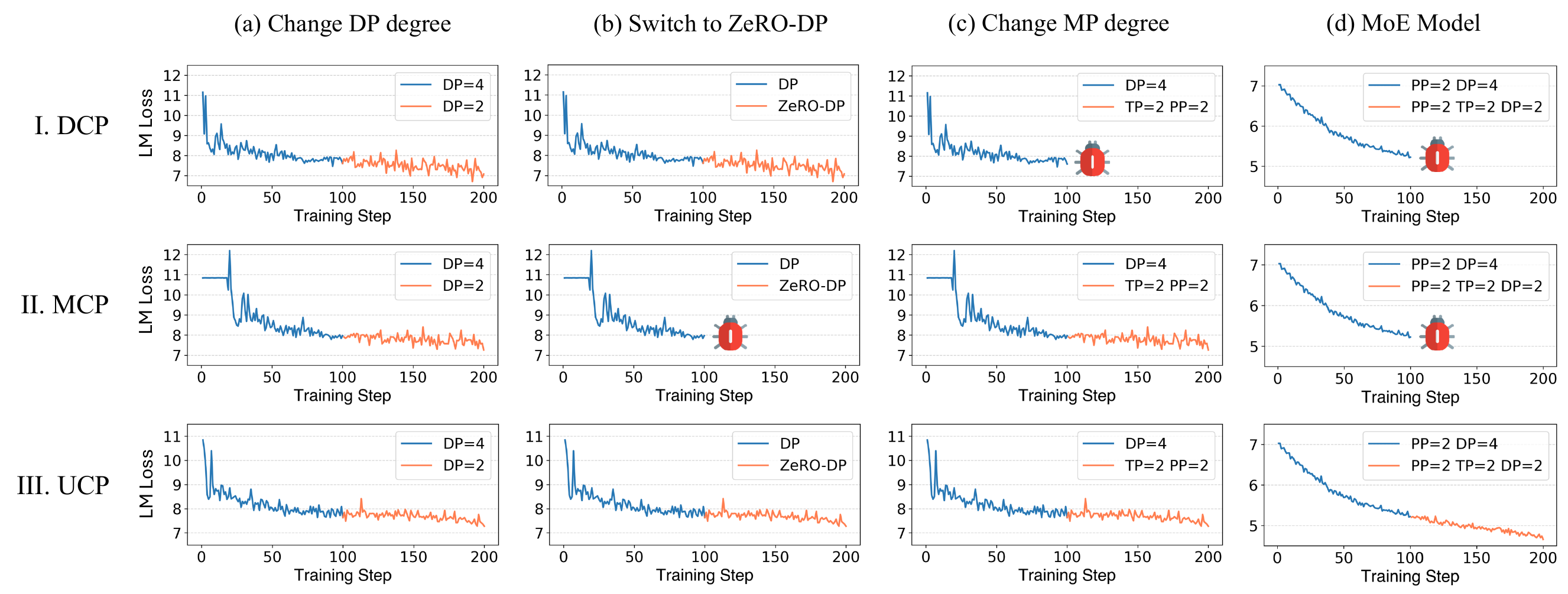}
\caption{Training curves of (a) changing DP degree, (b) switching to ZeRO-DP, (c) changing MP degree for GPT-3 medium, and (d) for a MoE model when resuming the training from I.DCP~\cite{dcp}, II. MCP~\cite{mcp}, III.UCP in the middle. "Bug" indicates errors encountered when changing parallelism strategy with a given checkpointing system.}
\label{fig:dcp}
\end{figure*}
\begin{figure*}[!ht]
\centering
\includegraphics[width=\textwidth]{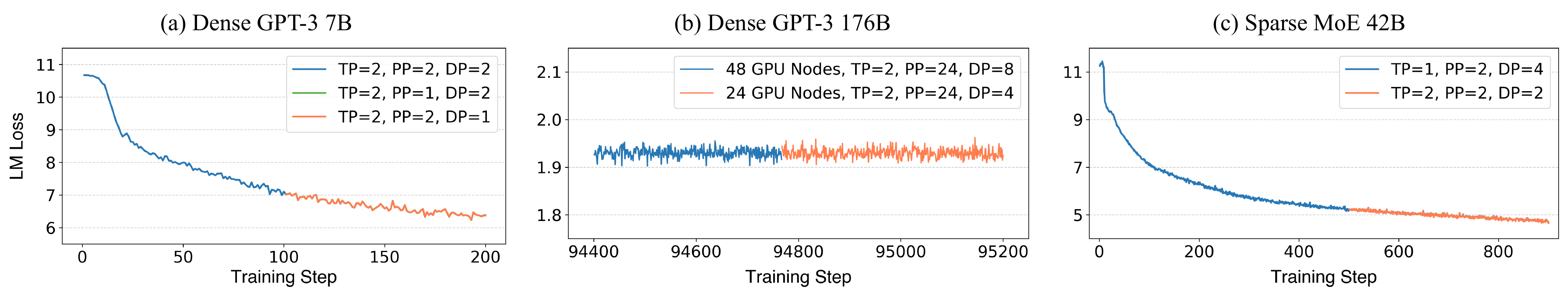}
\caption{Training curves of (a) {dense GPT-3 7B} (b) {dense GPT-3 176B} (c) {sparse MoE 42B} when resuming the training from \UCP in the middle.}
\vspace{-10pt}
\label{fig:correctness_arch}
\end{figure*}

\vspace{-2pt}
\paragraph{Multiple \emph{Source} to single \emph{Target}.} Fig.~\ref{fig:correctness}(b) shows the training curves from multiple \emph{Source} configurations to a single \emph{Target}. We fix all random seeds in the system so that all training job starts with the same model initialization states. We train the GPT-3 model using different \emph{Source} configurations. We then reconfigure their distributed checkpoints saved at the 100th iteration and resume training with a configuration of TP=2, PP=2, DP=1, and SP=1. The results show that regardless different \emph{Source} configurations, all training jobs can resume their training with the same \emph{Target} configuration. Most importantly, the resumed training curves match the curves from the \emph{Source} at iterations 101--200. These results validate the effectiveness of \UCP of reconfiguring parallelism while maintaining the training accuracy. 

Overall, \UCP is able to flexibly reconfigure parallelism strategies from any \emph{Source} and any \emph{Target} because its introduced atomic checkpoints decouple distributed checkpoints from specific parallelism and hardware configurations, which serves as a common representation for reconfigurable parallelism. Furthermore, the reconfiguration pipeline provides unified and automatic parallelism reconfiguration through tensor pattern matching and pattern-based transformation primitives for a wide range of parallelism strategies.

\subsection{Comparison with Alternative Methods}

As shown in Table~\ref{tab:comparison}, some checkpointing systems have expanded their capabilities to accommodate reconfigurable parallelism, such as DCP and MCP. To demonstrate the flexibility of state-of-the-art checkpointing systems and validate the results presented in Table~\ref{tab:comparison}, we compare the flexibility of DCP, MCP, and \UCP under four scenarios: (a) training with DP degree 4, then resume the training with DP degree 2; (b) switch to ZeRO-DP; (c) change the MP degree (including both PP and TP degree); (d) assessing flexibility when applied to a MoE model. For each scenario, we conduct training for 100 iterations, save checkpoints, then resume training with modified parallelism strategies for an additional 100 iterations.

As shown in Fig.~\ref{fig:dcp}, DCP supports changes in DP degree and switch to ZeRO-DP but fails to accommodate MP degree change. MCP, while supporting changing both DP and MP degree, shows limitations with MoE models and ZeRO-DP switch. 
It's important to note that "fail" and "limitation" in this context do not refer to an increase in training loss, but rather to system-level errors. In such cases, developers are required to either restart training from scratch or invest effort in crafting custom conversion scripts.
\UCP, on the other hand, successfully resumes training under all four scenarios. These results demonstrate \UCP's ability to reconfigure a larger and more complicated set of parallelism strategies for DL model training than baseline methods.


\subsection{Generalizability to Different Model Architectures}
\UCP is model architecture agnostic. As such, it is compatible with GPT models varying sizes and both dense and sparse model architectures. Fig.~\ref{fig:correctness_arch}
show the training convergence for GPT-3 7B~\cite{llama}, GPT-3 176B~\cite{bloom}, and a Mixtral-7x8B style MoE~\cite{mixtral}, when resuming from \UCP in the middle of training with new parallelism strategies. These figures show that training is seamlessly resumed with \UCP, achieving consistent convergence that aligns with the initial training phase across these diverse models. \UCP is able to achieve this because it considers a comprehensive set of parallelism strategies and model architectures when designing its tensor pattern matching and pattern-based transformation primitives, as demonstrated through the examples in \sref{subsubsec:pattern-based}. \revise{Notably, the GPT-3 176B curve corresponds to a real-world case during the BLOOM 176B~\cite{bloom} training. On July 4, 2022 at 9:00 PM, after nearly three month training, the allocation of 48 nodes expired, and the replacement cluster offered only 24 nodes. Using \UCP, the training job was seamlessly resumed on the smaller cluster without interruption. Without \UCP, training would have either halted prematurely—leaving tokens incompletely consumed and potentially degrading model quality—or required restarting the entire three-month training process from scratch.}




\subsection{Reconfiguration Efficiency}
\noindent
\textbf{Saving cost.}
As described in \sref{subsec:efficient-impl}, \UCP incurs no additional saving costs compared to existing distributed checkpointing mechanism and uses \emph{lazy reconfiguration invocation} to reconfigure parallelisms. The input of \UCP is the basic distributed checkpoint that is saved periodically. Therefore, the saving cost of \UCP is equivalent to that of the standard training process and does not impede the training speed. 

\begin{figure}[!t]
\centering
\includegraphics[width=\linewidth]{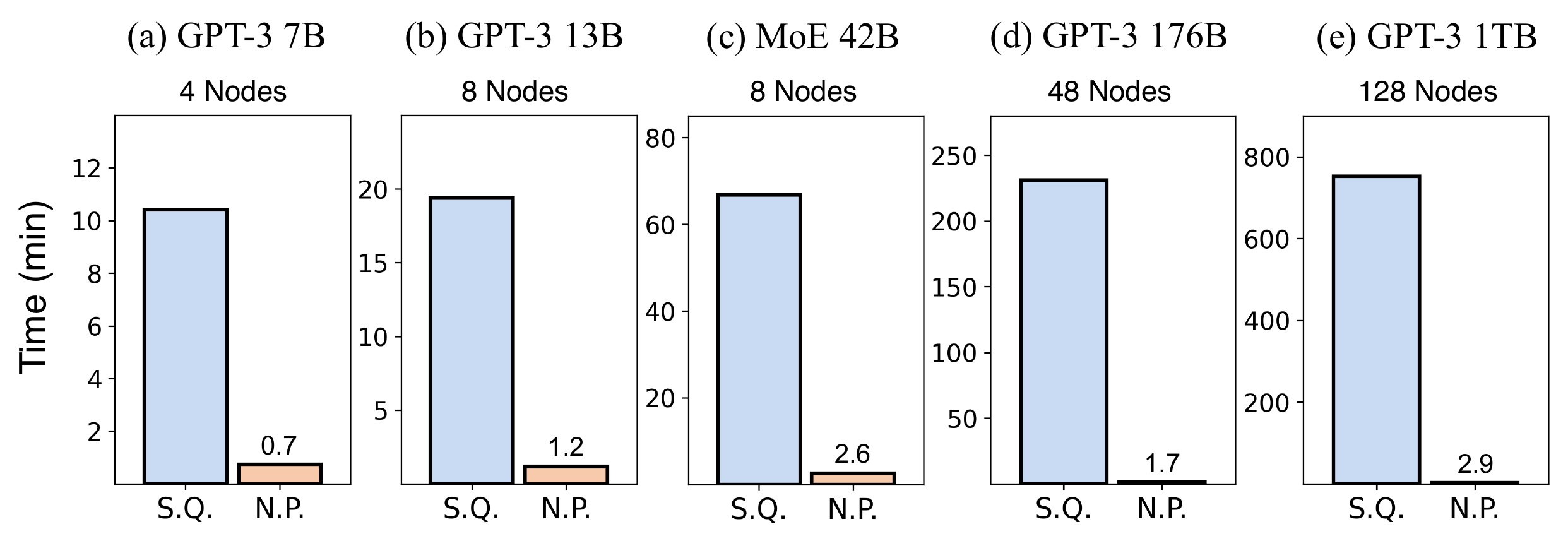}
\caption{Time cost for converting distributed checkpoints to \UCP atomic checkpoints with sequential (S.Q.) and nested parallel (N.P) approach across different model sizes.}
\vspace{-5pt}
\label{fig:conversion_time}
\end{figure}

\noindent
\textbf{Reconfiguration cost.}
The reconfiguration cost of \UCP involves both (1) the transformation overhead from distributed checkpoints to atomic checkpoints and (2) the loading of atomic checkpoints to all parallel workers. 
Fig.~\ref{fig:conversion_time} shows that the conversion time remains consistently bounded by approximately 3 minutes, even as model sizes scale up to 1T parameters. This efficiency is achieved through a combination of hardware scaling and the nested parallel conversion method (\sref{subsec:efficient-impl}).
The conversion can be executed directly within the training environment, which eliminates the need to transfer exceptionally large checkpoint files (up to TB in size) to different hardware infrastructure.
Compared to the sequential(S.Q.) conversion, \UCP nested parallelism(N.P.) approach achieves up to a 257x reduction in conversion time. Notably, when the model size increases, the number of GPU nodes also increases to partition the model to limited GPU memory, even for larger models such as LLaMA 3.1 405B that is trained on up to 2000 GPU nodes~\cite{llama3TechReport}, the conversion time will expected to be bounded to several minutes and does not increase linearly.

\begin{figure}[!ht]
\centering
\includegraphics[width=\linewidth]{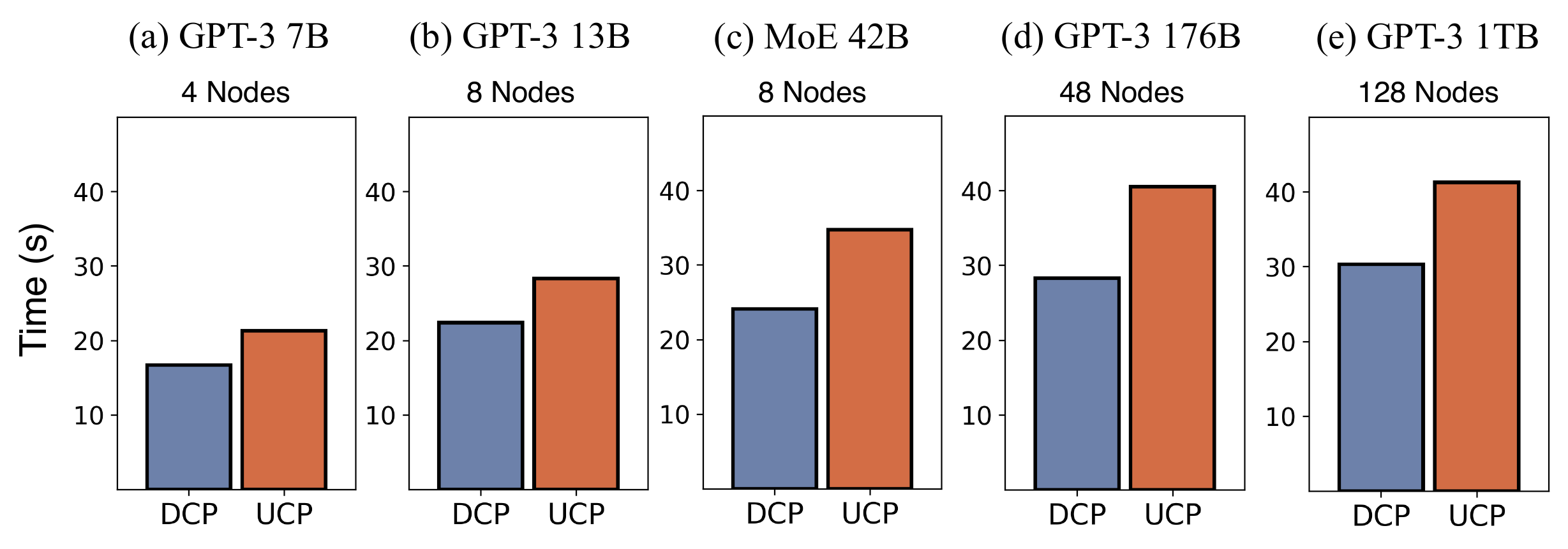}
\caption{Time cost for loading standard distributed checkpoints (DCP) and loading the UCP checkpoints across models in different sizes.}
\label{fig:loading_time}
\end{figure}

To measure the loading cost, we compare \UCP with directly loading from the saved distributed checkpoints. As standard distributed checkpoints cannot be loaded when there are changes in GPU counts or parallelism strategies, we keep the same GPU counts and parallelism strategies for the experiments.
As shown in \fref{fig:loading_time}, loading from atomic checkpoints incurs approximately 10s overhead compared to standard distributed checkpoints, consistently across different model size, because the loading time is primarily dominated by the checkpoint data volume, whereas \UCP does not change the checkpoint volume significantly.
\emph{Redundancy-bypassing loading} further eliminates redundancy when loading the atomic checkpoints within the same DP group. Moreover, the conversion and loading of atomic checkpoints is only triggered lazily (\sref{subsec:efficient-impl}). Therefore, \UCP's reconfiguration only accounts for a very small portion of the end-to-end training time.

\setlength{\tabcolsep}{13pt}
\begin{table*}[t]
\centering
\small
\begin{tabular}{ll|ccc|c}
\toprule
\textbf{Model} & \textbf{Hardware Config} & \textbf{Save (min)} & \textbf{Transform (min)} & \textbf{Load (min)} & \textbf{End-to-end (min)} \\
\midrule
GPT-3 7B      & 4 nodes A100           & 0.29 & 0.73 & 0.36 & \bf 1.38 \\
GPT-3 13B     & 8 nodes A100           & 0.38 & 1.17 & 0.47 & \bf 2.02 \\
MoE 42B       & 8 nodes A100           & 0.42 & 2.64 & 0.58 & \bf 3.64 \\
GPT-3 176B    & 48 nodes A100          & 0.48 & 1.67 & 0.68 & \bf 2.83 \\
GPT-3 1TB     & 128 nodes MI250X       & 0.50 & 2.93 & 0.69 & \bf 4.12 \\
\bottomrule
\end{tabular}
\vspace{-5pt}
\caption{End-to-end reconfiguration overhead for various models and hardware configurations.}
\vspace{-5pt}
\label{tab:reconfig-overhead}
\end{table*}

\revise{Table~\ref{tab:reconfig-overhead} summarizes the end-to-end reconfiguration cost, including checkpoint save, transformation, and load times. In summary, \UCP incorporates several system-level optimizations to minimize overhead. The Nested Parallel Reconfiguration is designed to minimize transformation costs, achieving up to a 257× reduction in conversion time. The Redundancy-bypassing loading optimizes loading time by avoiding redundant operations, resulting in a 3× to 20× speedup depending on the degree of data parallelism. The Lazy reconfiguration invocation reduces the frequency of invoking UCP, ensuring that it does not interfere with normal training. UCP is only triggered when a change in parallelism strategy is required.
Therefoer, even for models with 1T parameters, total reconfiguration remains under 5 minutes, which is negligible compared to training durations. Although these results are obtained using bandwidth levels comparable to consumer-grade SSDs, \UCP maintains functionality even under slower network and storage conditions. However, its performance in such scenarios will be limited by the available bandwidth and I/O throughput.}

\vspace{35pt}
\section{Related Work}
\label{sec:relate}

\noindent
\textbf{Checkpointing systems.} Checkpointing is widely used as the basic mechanism for fault recovery. While naive checkpointing stalls training, asynchronous checkpointing approaches enable the overlap of disk I/O and model operations~\cite{checkfreq,deepfreeze,veloc}. Just-In-Time (JIT)~\cite{jit} checkpointing employs the state redundancy from DP, creating checkpoints only after failures. To waive the bottleneck from the low bandwidth of remote persistent storage, FastPersist~\cite{fast-persist} achieves higher bandwidth by extending local NVMe SSDs, and in-memory checkpointing methods~\cite{gemini} utilize remote CPU memories, allowing checkpointing over high-speed training networks. These techniques primarily focus on optimizing checkpoint saving efficiency, which lies on the critical path of the overall training process. \UCP does not change the saving logic and is orthogonal to these optimization techniques, allowing for seamless integration.

\vspace{2pt}
\noindent
\textbf{Spot-instance training.} There has been research on using spot instances for DNN training as they are much more cost effective.
\revise{However, such dynamic environments make overhead of reconfiguration becomes a crucial factor. UCP already introduces several optimizations such as \textit{Nested Parallel Reconfiguration}, \textit{Redundancy-bypassing loading}, \textit{Lazy reconfiguration invocation}, that bring the reconfiguration cost down from hours to a few minutes, making it a potentially viable solution for more elastic training with spot instances. However, prior work, such as Bamboo~\cite{bamboo} and Parcae~\cite{parcae}, suggest that the dynamism in many cloud spot instances can be extremely high. Therefore, it may require more efficient parallelism reconfigurability to adapt to those environments. We believe that is a very interesting future research study.}

\vspace{2pt}
\noindent
\textbf{Reconfigurable ML systems.} VirtualFlow~\cite{virtualflow} and Singularity~\cite{singularity} decouple DL jobs from physical devices, which enable flexible GPU mapping at runtime but require complex driver-level virtualization and does not support multi-dimensional parallelism changes. Tenplex~\cite{tenplex} is a state management library by describing the state as a parallelizable tensor collection (PTC), and generates reconfiguration plans for job change at runtime. While Tenplex supports changing subsets of the parallelism, it is limited to intra-cluster reconfigurations, and storing partitioned parameters in host memory creates memory pressure and impacts training accuracy due to non-optimized state recovery. \UCP supports cross-cluster migration while fully preserving training accuracy.

\section{Conclusion}
\label{sec:conclusion}

This paper presents Universal Checkpointing (UCP), a checkpointing system for training large-scale DNNs with reconfigurable parallelism. The primary goal of \UCP is to enable flexible reconfiguration of parallelism strategies through distributed checkpoints. Therefore, \UCP provides a new checkpoint structure called atomic checkpoints and builds a pattern-based reconfiguration pipeline that enables automatic and systematic reconfiguration of a broad set of commonly used parallelism strategies while preserving model accuracy. To reduce the reconfiguration cost, \UCP introduces nested-parallel reconfiguration and lazy invocation to make the reconfiguration overhead negligible to the end-to-end training cost. 
Our evaluations on large-scale LLM training show that \UCP enables reconfiguration of a larger and more complex set of parallelism strategies with low cost. \UCP has been implemented and open-sourced via a PyTorch library for accelerating large-scale DNN training, and its effectiveness has been verified through real-world large-scale LLM pre-training tasks.  





\vspace{6pt}
\section*{Acknowledgments}

We sincerely appreciate the anonymous reviewers and our shepherd Young-ri Choi. Their insightful feedback helps significantly improve the quality of the paper. This research was supported by the National Science Foundation (NSF) under Grant No. 2441601. The work utilized the DeltaAI system at the National Center for Supercomputing Applications (NCSA) through allocation CIS240055 from the Advanced Cyberinfrastructure Coordination Ecosystem: Services \& Support (ACCESS) program, which is supported by National Science Foundation grants \#2138259, \#2138286, \#2138307, \#2137603, and \#2138296. The Delta advanced computing resource is a collaborative effort between the University of Illinois Urbana-Champaign and NCSA, supported by the NSF (award OAC 2005572) and the State of Illinois. This work also utilized the Illinois Campus Cluster and NCSA NFI Hydro cluster, both supported by the University of Illinois Urbana-Champaign and the University of Illinois System.

\vspace{6pt}
\section*{Availability}

Universal Checkpointing source code is available at \href{https://github.com/deepspeedai}{DeepSpeed}. For guidance on using and deploying Universal Checkpointing, you can refer to the \href{https://huggingface.co/docs/transformers/deepspeed}{Huggingface tutorial} and \href{https://github.com/deepspeedai/Megatron-DeepSpeed/tree/main/examples_deepspeed/universal_checkpointing}{Megatron-DeepSpeed examples}.

\newpage

\bibliographystyle{plain}
\bibliography{reference.bib}


\end{document}